\RequirePackage{fix-cm}
\documentclass[smallextended]{svjour3}       
\smartqed  
\usepackage[T1]{fontenc}
\usepackage[utf8]{inputenc}
\usepackage{xpatch}
\usepackage[usenames,dvipsnames]{xcolor}
\usepackage[english]{babel}
\usepackage[babel]{csquotes}
\usepackage{microtype}
\usepackage{ae,aecompl}
\usepackage{mathptmx}
\usepackage{mathrsfs}
\usepackage{amsfonts}
\usepackage{amsmath}
\usepackage{amsfonts}
\usepackage{mathtools}
\usepackage{graphicx}

\usepackage{float}
\usepackage{caption}
\usepackage{wrapfig}
\usepackage{hyperref}
\hypersetup{colorlinks=true,urlcolor=black,linkcolor=RoyalBlue,citecolor=Green,linktocpage=true}
\usepackage{tikz}
\usetikzlibrary{decorations.markings}
\usetikzlibrary{decorations.pathmorphing}
\newcommand{\nc}{\newcommand}
\nc{\be}{\begin{equation}}
\nc{\ee}{\end{equation}}
\nc{\ba}{\begin{eqnarray}}
\nc{\ea}{\end{eqnarray}}
\nc{\bea}{\begin{eqnarray}}
\nc{\eea}{\end{eqnarray}}
\nc{\nn}{\nonumber}

\begin{document}

\title{Gravitational collapse in Quantum Einstein Gravity \thanks{Proceedings based on the talk given by A.P.\ at  the workshop Lema\^{\i}tre, ``Black Holes, Spacetime Singularities and Gravitational Waves'', held at the Vatican observatory, May 9$^{\rm th}$ - 12$^{\rm th}$ 2017.}}

\author{Alfio Bonanno \and Benjamin Koch \and \\ Alessia Platania}
\authorrunning{A. Bonanno, B. Koch, A. Platania}


\institute{Alfio Bonanno \at
              INAF, Osservatorio Astrofisico di Catania, via S. Sofia 78, I-95123 Catania, Italy \\
              INFN,  Sezione di Catania,  via S. Sofia 64, I-95123, Catania, Italy. \\
              \email{alfio.bonanno@oact.inaf.it}           
			  \and
			  Benjamin Koch \at
              Instituto de F\'isica, Pontificia Universidad Cat\'olica de Chile,
Av. Vicuna Mackenna 4860, Santiago, Chile \\
              \email{bkoch@fis.puc.cl }
              \and
			  Alessia Platania \at
			  INAF, Osservatorio Astrofisico di Catania, via S. Sofia 78, I-95123 Catania, Italy \\
              INFN,  Sezione di Catania,  via S. Sofia 64, I-95123, Catania, Italy. \\
              Universit\`a degli Studi di Catania, via S. Sofia 63, I-95123 Catania, Italy \\
              \email{alessia.platania@oact.inaf.it}
}

\date{Received: date / Accepted: date}

\maketitle

\begin{abstract}
The existence of spacetime singularities is one of the biggest problems of nowadays physics. According to Penrose, each physical singularity should be covered by a ``cosmic censor'' which prevents any external observer from perceiving their existence. However, classical models describing the gravitational collapse usually results in strong curvature singularities, which can also remain ``naked'' for a finite amount of advanced time. This proceedings studies the modifications induced by Asymptotically Safe Gravity on the gravitational collapse of generic Vaidya spacetimes. It will be shown that, for any possible choice of the mass function, Quantum Gravity makes the internal singularity gravitationally weak, thus allowing a continuous extension of the spacetime beyond the singularity.

\end{abstract}

\section{Introduction}
\label{sec.1}

The existence of spacetime singularities indicates the failure of the classical description of gravity within the framework of General Relativity. In fact, the evolution of causal geodesics across a singularity is not uniquely determined: the spacetime is thus ``geodesically incomplete'' \cite{geroch1,geroch2} and General Relativity loses its predictive power as a classical, deterministic theory.

Spacetime singularities are a rather general feature of General Relativity \cite{1970HP}. According to the Cosmic Censorship Conjecture (CCC) spacetime singularities must be covered by an event horizon \cite{1969CC}, so that no signal from the singularity can reach the future null infinity. In the case of black holes, the central singularity and the event horizon are dynamically produced by the gravitational collapse of a sufficiently massive object. Depending on the initial conditions, the event horizon may also emerge after the formation of the central singularity, thus leaving the singularity ``naked'' for a finite amount of advanced time (see \cite{josb} for a complete review). The presence of {naked singularities} is allowed by several classical models and implies a violation of the Cosmic Censorship Hypothesis. On the other hand, the gravitational collapse involves ultra-high energy densities and in such extreme situations quantum gravitational effects may play a crucial role. 

Asymptotically Safe Gravity constitutes a promising approach to construct a predictive quantum theory for the gravitational interaction in the framework of quantum field theory. According to the Asymptotic Safety conjecture, quantization of gravity results in a renormalizable quantum field theory whose high-energy completion is defined by a non-gaussian fixed point (NGFP) which attracts the gravitational renormalization group flow in the ultraviolet limit. Asymptotic Safety makes use of non-perturbative functional renormalization group (FRG) methods. The resulting running gravitational couplings allow to incorporate the leading quantum effects in ordinary gravitational phenomena by ``RG-improving'' classical solutions. This method has been extensively used to study quantum-corrected black holes \cite{br00,ko14,lare,cope,torres14,torres14b}. In particular, the study of  the gravitational collapse in a quantum-corrected Vaidya-Kuroda-Papapetrou (VKP) model \cite{vaidya66,kuroda,papa} shows that the singularity becomes ``whimper'' when quantum gravity effects are taken into account \cite{BKP}.

In this proceedings we review the problem of black holes formation in the context of Asymptotic Safety. The RG-improvement of classical Vaidya spacetimes allows to relate the dynamics of the gravitational collapse to the renormalization group flow evolution. Quantum gravitational effects thus emerge dynamically and alter the way the event horizon (EH) forms during the collapse. It will be shown that the main results obtained in \cite{BKP} for a VKP model, remain true in the case of generic Vaidya spacetimes. In particular, the singularity arising from the gravitational collapse is always integrable, independently of the particular mass function.

The proceedings is organized as follows. In Sect. \ref{sec.2} we introduce the generalized Vaidya spacetimes and we briefly review the classical VKP model for the gravitational collapse. The general formalism to study the nature of the singularity will be introduced in Sect. \ref{sec.3}. In Sect. \ref{sec.4} we discuss a class of quantum-corrected Vaidya spacetimes. In particular, we shall study the outcome of the gravitational collapse in the case of a generic power-law mass function and we will analyze the singularity nature for general RG-improved Vaidya spacetimes. Finally, Sect. \ref{sec.5} summarizes our results.

\section{Generalized Vaidya spacetimes}
\label{sec.2}

In this section we review the basic formalism to globally characterize the outcome of the gravitational collapse in a generalized Vaidya spacetime.

The spacetime structure around a spherically symmetric object of mass $m$ can be described by the usual Schwarzschild metric. The study of dynamical phenomena such as gravitational collapse or black holes evaporation requires this mass to be a function of the time coordinate. Using the Eddington-Finkelstein coordinates, the line element is written as
\begin{equation}
ds^2=-f(r,v)\,dv^2+2\,dv\,dr+r^2\,d\Omega^2 \;\;. \label{metric}
\end{equation}
Here $v$ is the advanced time and $f(r,v)$ is a lapse function whose expression reads
\begin{equation} \label{clapseclapse}
f(r,v)=1-\frac{2 \,G_0 \,m(v)}{r}\;\;,
\end{equation}
$G_0$ being the Newton's constant. The metric \eqref{metric}, with the lapse function \eqref{clapseclapse} defines the so-called Vaidya spacetime \cite{vaidya51}. The latter is an exact solution of the Einstein field equations in presence of a Type II fluid with energy density \cite{vaidya51,vaidya66}
\be \label{classtho}
\rho(r,v)=\frac{\dot{m}(v)}{4\pi r^2}\;\;.
\ee
The Vaidya spacetime is part of a larger class of metrics, known as generalized Vaidya spacetimes \cite{1999WW}. This family of metrics is characterized by a more general lapse function
\begin{equation}
f(r,v)=1-\frac{2\,M(r,v)}{r} \;\;, \label{gvlapse}
\end{equation}
in which the {generalized mass function} $M(r,v)$ depends on both the advanced time and the radial coordinate. The generalized Vaidya spacetimes can be obtained as solutions of the Einstein field equations in presence of both Type-I and Type-II fluids. The corresponding  stress-energy tensor reads \cite{goswami,joshi} 
\be \label{Tmunu}
T_{\mu\nu}={\rho\,l_\mu l_\nu}+
{(\sigma+p)(l_\mu n_\nu + l_\nu n_\nu)+ p g_{\mu\nu}}
\ee
where 
\begin{align}
&\rho (r,v) = \frac{1}{4\pi G_0 r^2} \, \frac{\partial M(r,v)}{\partial v}\; , \\
&\sigma(r,v) = \frac{1}{4\pi G_0 r^2} \, \frac{\partial M(r,v)}{\partial r} \;  , \\
&p(r,v) = - \frac{1}{8\pi G_0 r} \, \frac{\partial^2 M(r,v)}{\partial r^2} \;,
\end{align}
and $n_\mu l^\mu = -1\,$, $\;l_\mu l^\mu=0$. In the standard Vaidya spacetime $M(r,v)\equiv G_0\,m(v)$, so that the stress-energy tensor \eqref{Tmunu} reduces to that of pure radiation. 

The Vaidya metric describes the spacetime around a spherical object of variable mass $m(v)$ and can be used to model the gravitational collapse of a massive star. The dynamics of the collapse is completely determined by the specific form of the mass function. A particularly important and simple model for the gravitational collapse is the Vaidya-Kuroda-Papapetrou (VKP) model \cite{vaidya66,kuroda,papa}. The latter was one of the first counterexamples to the Cosmic Censorship Conjecture. In the VKP model the mass function is parametrized as follows
\begin{equation}
m(v)=\begin{cases}0 & v<0 \\ \lambda v & 0\leq v < \bar{v} \\ \bar{m} & v\geq \bar{v}\end{cases}\label{VKPm}
\end{equation}
For $v<0$ the spacetime is flat and empty. At $v=0$ the radiation shells emanating from the collapsing star are injected and focused towards  $r=0$. The mass of the central object starts growing linearly $m(v)=\lambda v$ and the spacetime develops a singularity in $r=0$. The process continues up to $v=\bar{v}$, when the Vaidya spacetime converges to the standard Schwarzschild metric.

Studying the geodesic equation for outgoing null rays allows to determine the outcome of the collapse. In fact, if the singularity is not covered by an event horizon, there will be light rays reaching the future null infinity and terminating at the singularity in the past. In this case the singularity is \textit{globally naked}. Otherwise, the collapse results in a black hole.

In a generalized Vaidya spacetime the geodesic equation for outgoing light rays reads
\begin{equation}
\frac{dr}{dv} =\frac{1}{2}\left(1-\frac{2 M(r,v)}{r}\right) \label{vaidya} \,\,.
\end{equation}
In the special case of the VKP model, the above equation can be analytically solved and its general solution can be written as
\begin{equation}
-\frac {2\,\mathrm{ArcTan}\left[\frac {v - 4 r(v)} {v\sqrt {-1 + 16\,\lambda\,G_0}} \right]} {\sqrt {-1 +16\,\lambda\,G_0}}+\mathrm{log}\Big[ 2\lambda G_0v^2 -{r(v)}\,v+2\,{r(v)^2}\Big]=C \label{sol1}
\end{equation}
where $C$ is an arbitrary integration constant. The classical VKP model is characterized by a critical value of the radiation rate, $\lambda_c\equiv 1/16 G_0$, below which the singularity is {globally naked}. In this case the family of solutions in eq. \eqref{vaidya} reduces to the following implicit equation \cite{israel,joshiKP1,joshiKP2}
\begin{equation}
\frac{|r(v)-\mu_- v|^{\,\mu_-}}{|r(v)-\mu_+ v|^{\,\mu_+}}=\widetilde{C} \label{csol}
\end{equation}
where $\widetilde{C}$ is a complex constant and
\begin{equation}
\mu_\pm=\frac{1\pm\sqrt{1-16\, \lambda \,G_0}}{4} \;\;. \label{mupm}
\end{equation}
The implicit equation \eqref{csol} has two simple linear solutions $r_\pm(v)=\mu_\pm\, v$, with $\mu_\pm$ defined in \eqref{mupm}. The line $r_-(v)=\mu_-\, v$ is the tangent to the event horizon at $(r=0,v=0)$, while $r_+(v)=\mu_+\, v$ is the Cauchy horizon. The latter represents the first light ray emitted from the naked singularity. In particular, all geodesics lying between these two linear solutions in the $(r,v)$--plane are light rays starting from the singularity $(r=0,v=0)$ and reaching the observer at infinity. In this case the Cosmic Censorship Conjecture is violated in its weak formulation.

\section{Singularity structure in generalized Vaidya spacetimes}
\label{sec.3}

The most severe problem related to the existence of singular spacetimes is the impossibility of uniquely determining the evolution of the spacetime beyond the singularity. According to the singularity theorems \cite{1970HP} the existence of singularities in the solutions of Einstein field equations is quite general. However, such theorems do not specify the properties of such singularities, in particular their ``nature''. 

The physical relevance of a singularity is determined by its strength \cite{josb,goswami}. Following the Tipler classification \cite{tipler77}, a singularity is ``strong'' if an object falling into the singularity is destroyed by the gravitational tidal forces, thus disappearing from the spacetime once the singularity is reached. According to \cite{tipler77}, only strong curvature singularities are physically relevant. On the contrary the gravitationally weak or ``integrable'' singularities are considered to be less severe, as the spacetime may be continuously extended across the singularity (see \cite{joshi} for an extensive discussion).

The singularity strength is determined by the behavior of light-like geodesics in the vicinity of the singularity \cite{goswami}. Let us consider a generalized Vaidya spacetime with mass function $M(r,v)$. The geodesic equation for null rays is conveniently cast in the form of a system of coupled first order differential equations
\begin{equation}
\begin{cases}\frac{{d}v(t)}{{d}t}=N(r,v)\equiv2\,r \\
\frac{{d}r(t)}{{d}t}=D(r,v)\equiv r-2\,M(r,v) \;\;.\end{cases} \label{ds}
\end{equation}
The fixed point solutions of the system \eqref{ds} are identified by the conditions $r=0$ and $M(0,v)=0$. They coincide with the singular loci of the generalized Vaidya spacetimes \cite{goswami}. Linearizing the system around a fixed point solution $(r_\mathrm{FP},v_\mathrm{FP})$ yields 
\begin{equation}\label{dslin}
\begin{cases}\frac{\mathrm{d}v(t)}{\mathrm{d}t}=\dot{N}_\mathrm{FP}\,(v-v_\mathrm{FP})+N'_\mathrm{FP}\,(r-r_\mathrm{FP}) 
\\ \frac{\mathrm{d}r(t)}{\mathrm{d}t}=\dot{D}_\mathrm{FP}\,(v-v_\mathrm{FP})+D'_\mathrm{FP}\,(r-r_\mathrm{FP})\end{cases} 
\end{equation}
where a ``prime'' denotes differentiation respect to $r$, a ``dot'' stands for differentiation with respect to $v$, and the subscript FP means that the derivatives are evaluated at the fixed point. The linearized system \eqref{dslin} defines the Jacobian matrix $J$
\be
J\equiv %
\begin{pmatrix}
\dot{N}_\mathrm{FP} & N'_\mathrm{FP} \\
\dot{D}_\mathrm{FP} & D'_\mathrm{FP}
\end{pmatrix} \;\;.
\ee
In this description the behavior of the light-like geodesics in proximity to a fixed point is completely determined by the eigenvalues of $J$
\begin{equation}
\gamma_\pm=\frac{1}{2}\left(\mathrm{Tr}J\pm\sqrt{(\mathrm{Tr}J)^2-4\,\mathrm{det}J}\right) \;, \label{valueeig}
\end{equation}
where
\begin{subequations} \label{dyntrdet}
\begin{align}
&\mathrm{Tr}J=\dot{N}_\mathrm{FP}+D'_\mathrm{FP}\equiv1-2\,M'_\mathrm{FP} \\
&\mathrm{det}J=\dot{N}_\mathrm{FP}D'_\mathrm{FP}-\dot{D}_\mathrm{FP}N'_\mathrm{FP}\equiv4\,\dot{M}_\mathrm{FP} \;\;,
\end{align}
\end{subequations}
and by the corresponding eigendirections. The latter define the characteristic lines
\begin{equation}
r_\pm(v)=r_\mathrm{FP}+\frac{\gamma_\pm}{2}\,(v-v_\mathrm{FP}) \label{lleig}
\end{equation}
passing through the fixed point $(r_\mathrm{FP},v_\mathrm{FP})$. The slope of these lines determines the way radial null geodesics approach the fixed point. This description allows to locally characterize the singularity and determine its strength.

The central singularity is \textit{locally naked} if there are light-like geodesics starting from the singularity with a well defined tangent vector $\mathbf{v}$ and reaching the observer at infinity. In terms of dynamical systems, a well defined tangent vector exists when the conditions $\mathrm{det}J>0$ and $(\mathrm{Tr}J)^2-4\,\mathrm{det}J>0$ are simultaneously satisfied. Moreover, as we are considering outgoing light ray, the additional constraint $\mathrm{Tr}J>0$ is required. A locally naked singularity is thereby a \textit{repulsive node} of the dynamical system \eqref{ds}. The slope $X_\mathrm{FP}$ of the tangent vector $\mathbf{v}$
\be
\displaystyle{X_\mathrm{FP}\equiv \lim_{(r,v)\to \mathrm{FP}}\tfrac{v(r)}{r}}
\ee
is then determined by the non-marginal eigendirection \eqref{lleig} tangent to light-like geodesics at the singularity. Denoting by $\bar\gamma$ the eigenvalue \eqref{valueeig} associated to this eigendirection, one easily finds $X_\mathrm{FP}={2}/{\bar\gamma}$. Finally, following \cite{goswami}, the parameter
\begin{equation}
S=\tfrac{X_\mathrm{FP}^2}{2}\,(\partial_v M)_\mathrm{FP} \label{strengthparameter}
\end{equation}
is a measure of the singularity strength. A strong curvature singularity is  identified by the condition $S>0$. On the contrary, if $S\leq0$, the singularity is gravitationally weak (or integrable). When applying this analysis to the classical VKP model, one finds that the central singularity, represented by the fixed point $(0,0)$ in the $(r,v)$--plane, is always strong \cite{goswami}. Moreover the classical critical value $\lambda_c=1/16G_0$ already emerges at the level of the linearized system: when $\lambda>1/16G_0$ the fixed point is a spiral node. Otherwise $(0,0)$ is a pure repulsive node and corresponds to a locally naked singularity \cite{israel}.

\section{RG-improved Vaidya spacetimes}
\label{sec.4}

In this section we study the modifications of the classical Vaidya spacetimes induced by quantum gravitational effects in the framework of Asymptotically Safe Gravity.

The Asymptotic Safety scenario for Quantum Gravity relies on FRG techniques. The gravitational effective average action $\Gamma_k$ is a function of the renormalization group (energy) scale $k$. The RG evolution of $\Gamma_k$ through the theory space is dictated by the Wetterich equation \cite{1993W,1994M,ReuterWetterich,martin}
\begin{equation} \label{wett}
k\partial_k \Gamma_k=\frac{1}{2}\mathrm{Tr}\left(\frac{k\partial_k \mathrm{R}_k}{\Gamma_k^{(2)}+\mathrm{R}_k}\right)
\end{equation}
Once an ansatz for the physical action has been chosen, eq. \eqref{wett} allows to derive the beta functions $\beta_{g_i}(\bold{g})$ for the running couplings $\bold{g}\equiv(g_1,g_2,\dots,g_N)$ under consideration. The RG fixed points are then determined by the conditions $\beta_{g_i}(\bold{g}_\ast)=0$. Using the FRG one can study the behavior of the running functions $g_i(k)$ with the energy scale $k$. In particular, the running Newton's coupling can be approximated by \cite{br00}
\begin{equation}\label{running}
G_k=\frac{G_0}{1+\omega \, G_0 \, k^2}
\end{equation}
where $\omega=1/g_\ast$, $g_\ast>0$ being the value of the dimensionless Newton's constant at the NGFP.

We now employ a RG improvement procedure to study the Quantum Gravity effects in the gravitational collapse of a massive spherical star. Following \cite{br00,BKP}, we start from the classical Vadya solution \eqref{metric} and perform the RG improvement by replacing $G_0$ with the running Newton's coupling $G_k$, eq. \eqref{running}. The RG-improved lapse function thus reads
\begin{equation} \label{lapseimp}
f_\mathrm{I} (r,v)=1-\frac{2 \,m(v)}{r}\,\,\frac{G_0}{1+\omega \, G_0 \, k^2}\;\;.
\end{equation}
The inclusion of quantum corrections directly modifies the spacetime geometry \eqref{metric}. This strategy is thus appropriate for studying quantum-corrected black holes solutions \cite{1999alfiomartin,br00,2006alfiomartin,ko14,torres15,BKP} in the context of Asymptotic Safety.

A consistent description of the RG-improved gravitational collapse requires to find a relation between the renormalization group scale $k$ and the actual collapse dynamics. Since the formation of realistic black holes is caused by the gravitational collapse of both matter and radiation, one can argue that the energy density of the collapsing fluid may serve as a physical infrared cutoff. Therefore, we shall use the following scale-setting
\be
\label{ci}
k \equiv \xi \sqrt[4]{\rho}
\ee
where $\xi$ is an arbitrary positive constant and the specific functional form $k(\rho)$ is dictated by simple dimensional arguments (see also \cite{BKP}). At last, since $\rho$ is the energy density of the classical (bare) inflowing radiation, eq.~\eqref{classtho}, the infrared cutoff $k$ reads
\be
k(r,v)\equiv\xi\sqrt[4]{\frac{\dot{m}(v)}{4\pi r^2}} \;\;. \label{cutidBH}
\ee
Combining the RG-improved lapse function \eqref{lapseimp} with the above cutoff identification finally yields 
\begin{equation}
f_\mathrm{I}(r,v)=1-\frac{2\,m(v)\,{G_0}}{r+\alpha \, \sqrt{\dot{m}(v)}}\;, \label{qlapse}
\end{equation}
where $\alpha={\omega \,\xi^2 \,G_0}/{\sqrt{4\pi}}$. The zeros of the lapse function defines the so-called apparent horizon (AH). In our case the AH reads
\begin{equation}
r_\mathrm{AH}(v)=2\,m(v)\,G_0-\alpha\,\sqrt{\dot{m}(v)}\;\;,
\end{equation}
with the condition $r_\mathrm{AH}(v)\geq0$. When the collapse ends, the RG-improved Vaidya model defined by eq. \eqref{qlapse} reduces to a quantum-corrected Schwarzschild metric with lapse function
\be
f_S(r,v)=1-\frac{2\,m(\bar{v})\,{G_0}}{r+\alpha \, \sqrt{\dot{m}(\bar{v})}} \;\;.
\ee
The latter corresponds to a Schwarzchild black hole with 
mass $\bar{m}={m}(\bar{v})$. In this case the apparent and event horizons coincide and define a RG-improved Schwarzchild radius $r_S= 2\,{m}(\bar{v})\,G_0-\alpha\sqrt{\dot{m}(\bar{v})}$. 

Remarkably, provided $\dot{m}\neq0$, the lapse function $f_\mathrm{I}(r,v)$ is regular in $r=0$. Nevertheless, the Ricci scalar $R$ and Kretschmann scalar $ K=R_{\alpha\beta\gamma\delta}R^{\alpha\beta\gamma\delta}$
\begin{subequations}
\begin{align}
&R=\frac{4\alpha^2 m(v) \dot{m}(v)} {r^2\big(r + \alpha\sqrt {\dot{m}(v)} \;\big)^3} \;\;, \\[0.3cm] 
&K=\frac {16\, m(v)^2\left(r^4 + {\big(r + \alpha\sqrt {\dot{m}(v)} \big)^2} r^2 +{\big (r + \alpha\sqrt{\dot{m}(v)} \big)^4}  \right)} {r^4 \big (r + \alpha\sqrt {\dot{m}(v)} \big)^6} \;\;,
\end{align}
\end{subequations}
diverge as $r\to0$ and hence the hypersurface $r=0$ is singular.

\subsection{Gravitational collapse in the RG-improved Vaidya spacetime}
\label{sec.333}

The causal structure of the RG-improved spacetime can be studied in the standard way, namely by analyzing the solutions to the geodesic equation for outgoing radial null rays. The RG-improved version of the geodesic equation reads
\begin{equation}
\dot{r}(v)=\frac{1}{2}\left(1-\frac{2\,m(v)\,G_0}{r(v)+\alpha \, \sqrt{\dot{m}({v})}}\right) \;\;. \label{eqi}
\end{equation}
We now specialize our discussion to the case $m(v)=\lambda v^n$. In the special case $n=1$ (VKP model), eq. \eqref{eqi} can be solved analytically and its general solution reads \cite{BKP}
\begin{equation} \label{eqgeoBKP}
\mathrm{log}\Big[ 2\lambda G_0 v^2- {(r(v)+\alpha \sqrt{\lambda})}\,{v}+2\,{(r(v)+\alpha\sqrt{\lambda})^2}\Big]-\frac {2\mathrm{ArcTan}\left[\frac {v - 4 \, (\,r(v)+\alpha \sqrt{\lambda}\,)} {v\sqrt {-1 + 16\,\lambda\,G_0}} \right]}{\sqrt {-1 +16\,\lambda\,G_0}}=C
\end{equation}
where $C$ is an integration constant. 

The appearance of a naked singularity depends on whether the event horizon (EH) ``emerges'' before or after the formation of the singularity in $r=0$. The event horizon can be seen as a curve $r_\mathrm{EH}(v)$ in the $(r,v)$--plane which separates outgoing null geodesics converging back to singularity from those reaching the future null infinity. As the singularity forms at $v=0$, if $r_\mathrm{EH}(0)\leq0$ the singularity is initially naked.

\begin{figure}
\centering
\includegraphics[width=0.7\textwidth]{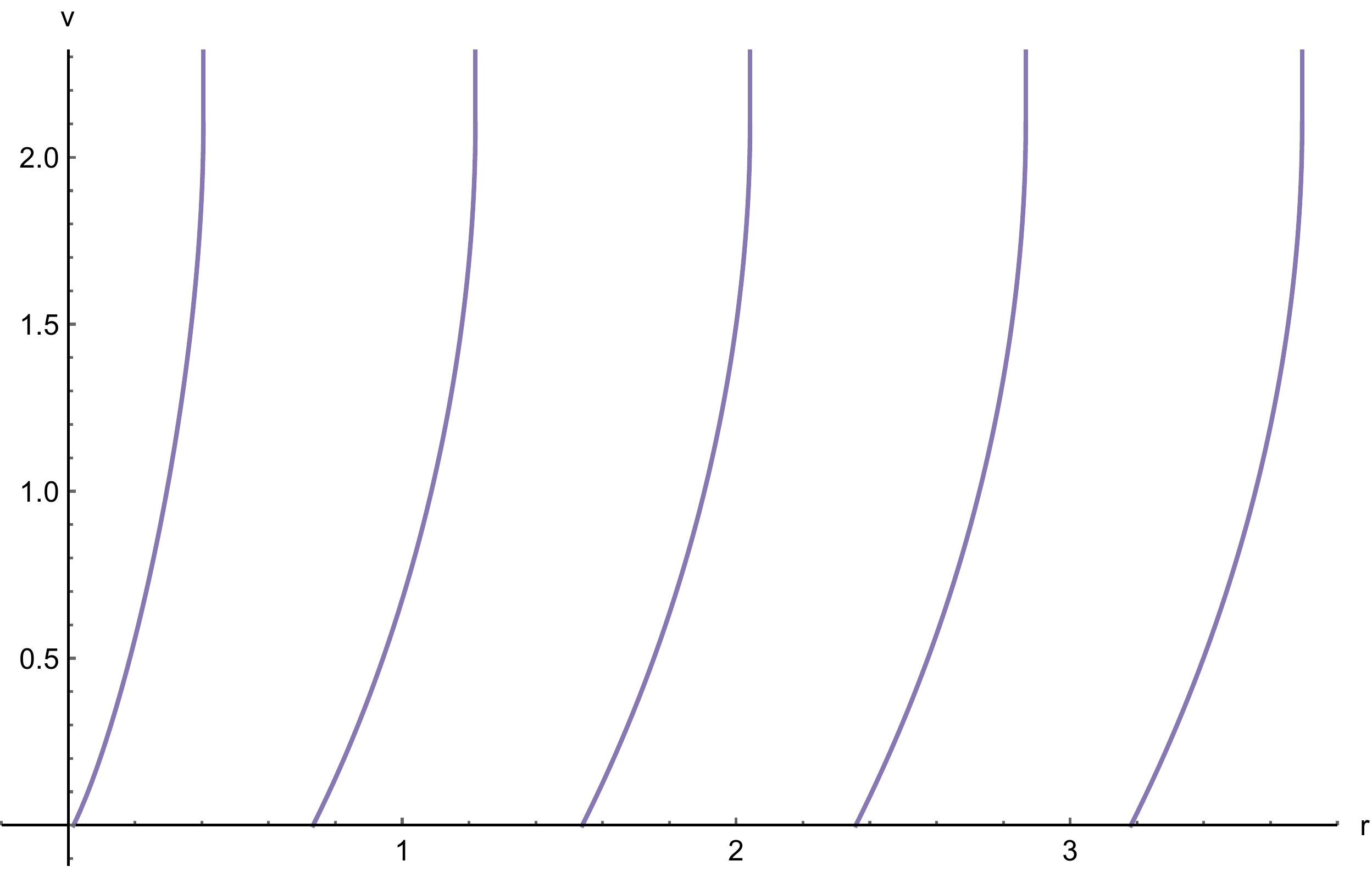}
\caption{Event horizon $r_\mathrm{EH}(v;\lambda)$ of the VKP model for various values of the radiation rate $\lambda$ and $\bar{v}=2$. The initial radius of the EH, $r_\mathrm{EH}(0;\lambda)$, is a monotonically increasing function of $\lambda$. The critical value below which the singularity is naked is identified by the limiting condition $r_\mathrm{EH}(0;\lambda_c)=0$.}
\label{EH1}
\end{figure}

\begin{figure}
\centering
\includegraphics[width=0.6\textwidth]{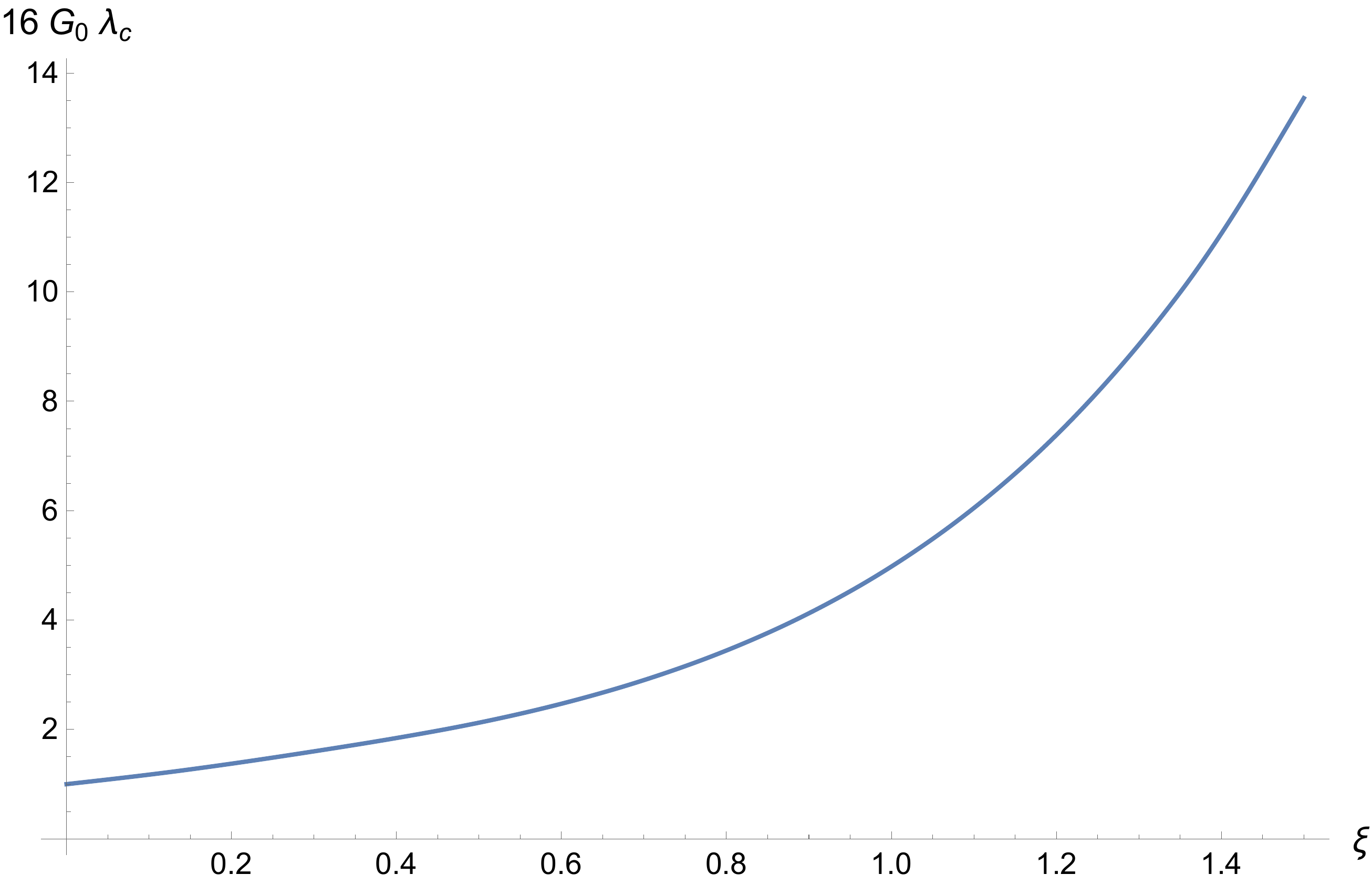}
\caption{Relative critical value ${16 G_0}\lambda_c$ as a function of $\xi$ in the RG-improved VKP model \cite{BKP}. The critical value $\lambda_c$ is always greater than the classical one, which is recovered at $\xi=0$.}
\label{figxi}
\end{figure}

\begin{figure}
\centering
\includegraphics[width=0.6\textwidth]{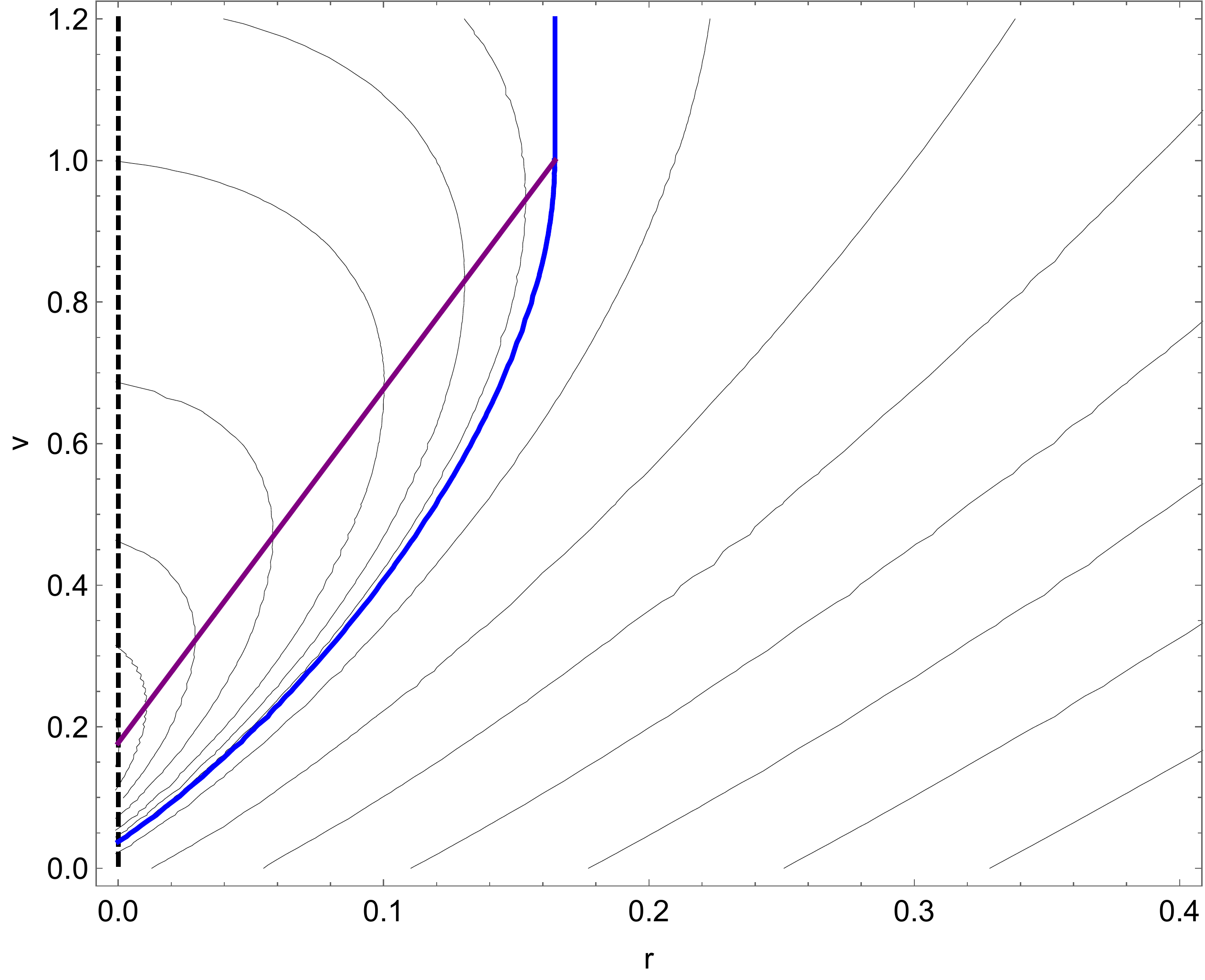}
\caption{Global behavior of light-like geodesics in the $(r,v)$--plane in the RG-improved VKP model. This phase diagram has been obtained in \cite{BKP} for $\lambda\leq\lambda_c$. The blue line is the EH, the purple line is the AH, and the black curves are obtained from the implicit equation \eqref{eqgeoBKP} by varying the constant $C$. The EH forms after the appearance of the central singularity $r=0$. The singularity remains (globally) naked for a finite amount of advanced time.}
\label{figq}
\end{figure}

Fig.~\ref{EH1} shows the shape of the event horizon $r_\mathrm{EH}(v;\lambda)$ for different values of $\lambda$, $n=1$ and $\bar{v}=2$. By lowering the radiation rate $\lambda$, the initial value $r_\mathrm{EH}(0;\lambda)$ decreases. The value of $\lambda$ such that $r_\mathrm{EH}(0;\lambda_c)=0$ identifies the critical value $\lambda_c$ below which the singularity generated by the collapse is globally naked. In general, the critical value $\lambda_c$ cannot be analytically determined. In \cite{BKP} it has been shown that the critical value arising in a RG-improved VKP model is always greater than the classical one, as it is apparent from Fig. \ref{figxi}.
The behavior of light-like geodesic in the case $\lambda\leq\lambda_c$ is described by the implicit equation~\eqref{eqgeoBKP} and the corresponding phase diagram, obtained in \cite{BKP}, is shown in Fig. \ref{figq}.

\begin{figure}
\centering
\includegraphics[width=0.8\textwidth]{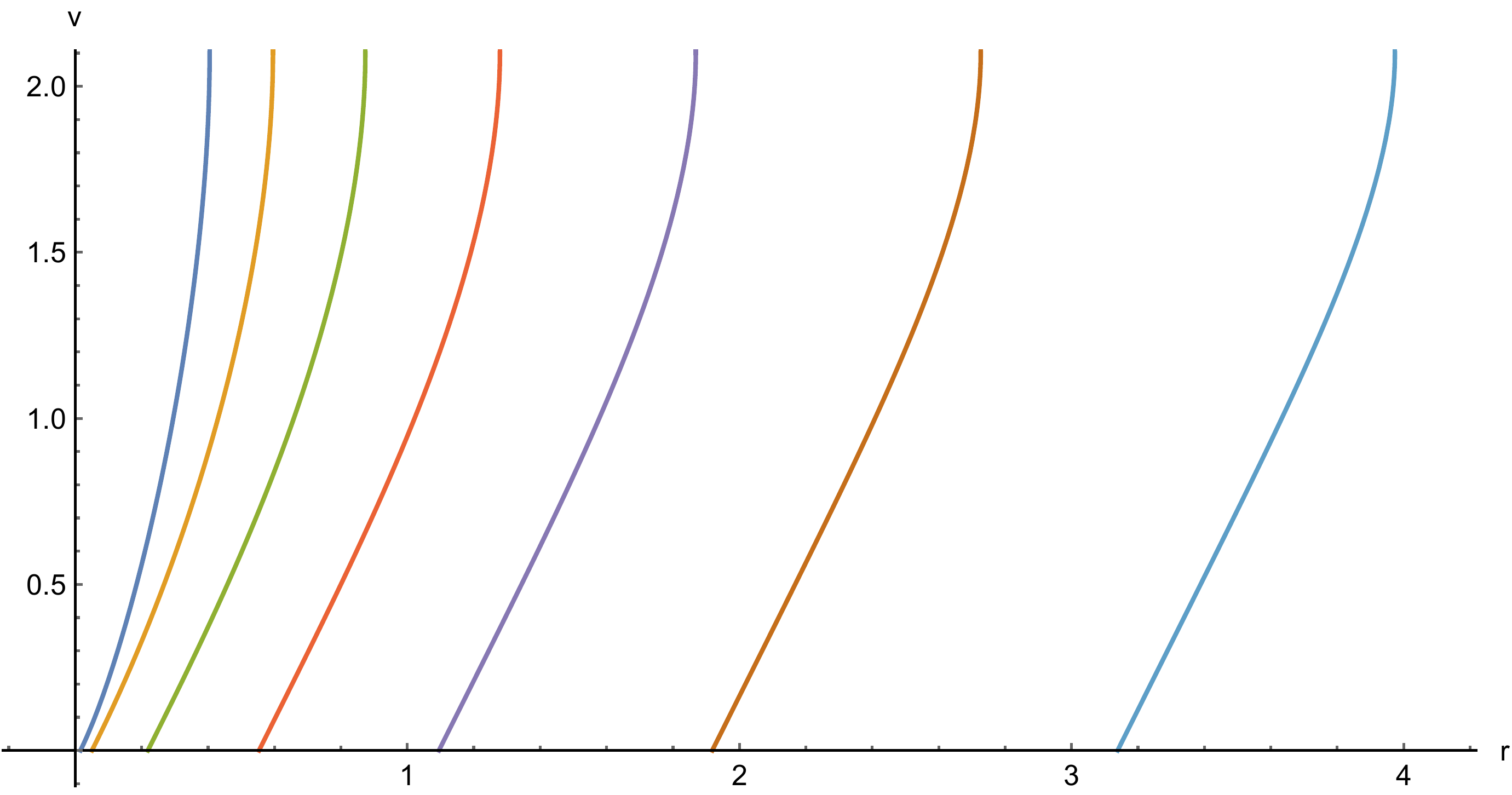}
\caption{Event horizon $r_\mathrm{EH}(v;n)$ produced by the mass function $m(v)=\lambda v^n$ for various values of $n$, $\bar{v}=2$ and $\lambda\equiv\lambda_c$, with $\lambda_c$ the critical value for $n=1$. The innermost EH corresponds to the case $n=1$ (VKP model). The initial  radius of the event horizon, $r_\mathrm{EH}(0;n)$, increases with $n$. As a consequence, the bigger is $n$ the lower is the corresponding critical value $\lambda_c^{(n)}$. This suggests that sufficiently rapid gravitational collapses avoid the formation of naked singularities.}
\label{EH2}
\end{figure}

\begin{figure}
\centering
\includegraphics[width=0.75\textwidth]{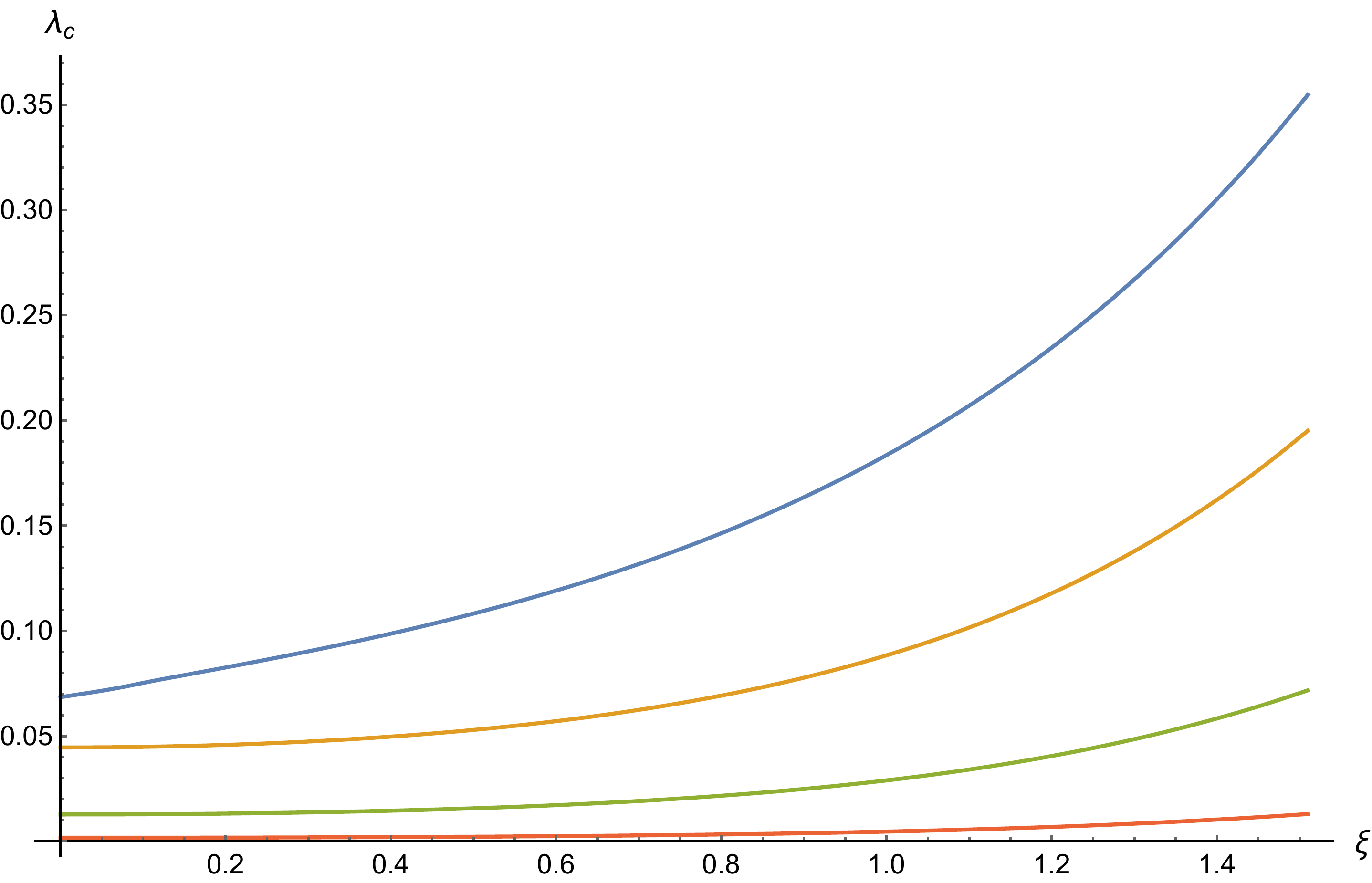}
\caption{Critical value $\lambda_c(\xi)$ associated with the mass function $m(v)=\lambda v^n$ for various values of $n$ and $\bar{v}=2$. The blue, yellow, green and orange curves correspond to $n=1,3,5,8$ respectively. From this picture is evident that, independently of $\xi$, high values of $n$ provide low values for $\lambda_c$. Naked singularities are thus disfavored when the collapse is sufficiently rapid.}
\label{xin}
\end{figure}

In the case $m(v)=\lambda v^n$ the shape of the event horizon, the critical value $\lambda_c$ and the formation of naked singularities strongly depend on the power index $n$. In particular, for a fixed value of $\lambda$, the quantity $r_\mathrm{EH}(0;n)$ increases with $n$ (see  Fig.~\ref{EH2}). This entails that the corresponding critical value $\lambda_c^{(n)}$ diminishes as $n$ increases (see Fig.~\ref{xin}).  Naked singularities are thus disfavored when the collapse is sufficiently rapid. 

Although quantum gravitational effects cause an increase of the critical value $\lambda_c$, thus favoring the occurrence of naked singularities, 
a running Newton's coupling vanishing in the ultraviolet limit turns the singularity of the VKP model into a line of integrable singularities \cite{BKP}. As we shall see in the next section, this fact remains true for a generic mass function $m(v)$. 


\subsection{Singularity structure in RG-improved Vaidya spacetimes}

We now apply the analysis described in Sect.~\ref{sec.3} to study the singularity structure in the quantum-corrected Vaidya model. 

The RG-improved Vaidya metric can be seen as a generalized Vadya spacetime with generalized mass function
\begin{equation} \label{quantumass}
M_I(r,v)=G(r,v)\,m(v)=\frac{G_0\,r}{r+\alpha\sqrt{\dot{m}(v)}} \,m(v)
\end{equation}
and corresponds to a non-trivial fluids mixture. If $\dot{m}\neq0$, the running Newton's coupling vanishes for $r\to0$, and the fixed point condition $M(0,v)=0$ holds at all values of the advanced time $v$. The singularity thus extends along the entire hypersurface $r=0$ and defines a line of fixed points. In general a line of fixed points arises when the Jacobian determinant is zero, $\mathrm{det}J=0$. In our case the latter condition is satisfied because the Newton's coupling vanishes in the ultraviolet limit
\be \label{Grvlim}
\mathrm{det}J\propto (\partial_v M)_\mathrm{FP}= \lim_{r\to0} \left(\dot{m}(v)-\frac{\alpha\,m(v)\,\ddot{m}(v)}{2 \sqrt{\dot{m}(v)} (r+\alpha\sqrt{\dot{m}(v)})}\right)\,G(r,v)=0 \;\;.
\ee
Moreover, the antiscreening behavior of the Newton's constant implies
\begin{equation}
S\propto (\partial_v M)_\mathrm{FP}\propto  \lim_{r\to0} G(r,v)=0 \;\;.
\end{equation}
Quantum Gravity fluctuations near the singularity convert the classical strong curvature singularity into a line of gravitationally weak singularities. This result does not depend on the particular cutoff identification \cite{BKP} and, according to eq.~\eqref{Grvlim}, the integrability of the singularity applies to all possible mass functions $m(v)$ such that $\dot{m}\neq0$.

\section{Conclusions}
\label{sec.5}

In this proceedings we reviewed the construction of a quantum-corrected Vaidya model \cite{BKP} for the gravitational collapse. The latter is obtained by incorporating the running of Newton's coupling, as predicted by Asymptotically Safe Gravity \cite{br00}, into the lapse function characterizing classical Vaidya spacetimes. We then used the resulting model to analyze the outcome of the gravitational collapse in the case of power law mass functions. This study showed that, at a quantum level, naked singularities are highly disfavored when the collapse is sufficiently rapid, as in the classical case. Finally, we generalized the study of the singularity nature performed in \cite{BKP} to the case of RG-improved Vaidya models with generic mass functions. In our model the strength parameter is always proportional to the value of Newton's constant at the singularity, independently of the specific form of the mass function. Since the running Newton's coupling vanishes in the ultraviolet limit \cite{br00}, this entails that the central singularity is gravitationally weak for any possible choice of the mass function.

Our model has been obtained through the first step of an iterative RG-improvement procedure. In order to construct a self-consistent quantum-Vaidya solution it would be necessary to study the next-to-leading order quantum modifications to the Vaidya model. We hope to address this issue in a future work.

\begin{acknowledgements}
We would like to thank the organizers of the workshop Lema\^{\i}tre
for their hospitality and for creating a highly stimulating scientific atmosphere. 
B.K. acknowledges Fondecyt 1161150.
\end{acknowledgements}


\bibliography{tsbib}
\bibliographystyle{spphys}  

\end{document}